\documentclass[12pt]{article}
\usepackage{amsmath}
\usepackage{epsfig}

\newcommand{\clh}{{\mathcal{H}}}
\newcommand{\clp}{{\mathcal{P}}}
\newcommand{\cle}{{\mathcal{E}}}
\newcommand{\clb}{{\mathcal{B}}}

\newcommand{\cll}{{\mathcal{L}}}
\newcommand{\clx}{{\mathcal{X}}}
\newcommand{\clz}{{\mathcal{Z}}}

\newcommand{\sigt}{\sigma_{\mbox{\scriptsize T}}}
\newcommand{\curl}{\,\mbox{curl}\,}
\newcommand{\kmsmpc}{\mbox{$\mathrm{km\,s^{-1}\,Mpc^{-1}}$}}

\title{\bf The Covariant Perturbative Approach to Cosmic Microwave
Background Anisotropies}
\author{Anthony Challinor\thanks{A.D.Challinor@mrao.cam.ac.uk}\\
{\it Astrophysics Group, Cavendish Laboratory, Madingley Road,} \\
{\it Cambridge CB3 0HE, UK.}}
\date{}

\begin{document}
\maketitle
\vspace{-\baselineskip} 

\begin{abstract}
The Ehlers-Ellis 1+3 formulation of covariant hydrodynamics, when supplemented
with covariant radiative transport theory, gives an exact, physically
transparent description of the physics of the cosmic microwave background
radiation (CMB). Linearisation around a Friedmann-Robertson-Walker (FRW)
universe provides a very direct and seamless route through to the linear,
gauge-invariant perturbation equations for scalar, vector and tensor modes in
an almost-FRW model. In this \emph{contribution} we review covariant
radiative transport theory and its application to the perturbative method
for calculating and understanding the anisotropy of the
CMB. Particular emphasis is placed on the inclusion of polarization in
a fully covariant manner. With this inclusion, the covariant
perturbative approach offers a complete description of linearised CMB physics
in an almost-FRW universe.
\end{abstract}

\section{Introduction}
\label{intro}

Given the potential impact of observations of the
cosmic microwave background radiation on our understanding of the
large scale properties of the universe, there is a strong
case for developing a physically transparent and portable formalism for the
study of CMB physics. Ideally, such a framework should allow one to give:
(i) an exact, but physically illuminating description of all the relevant
physics;
(ii) a simple, gauge-invariant linear perturbation theory around a variety
of background models. Along with many other workers in this
\nocite{stoeger95b,maartens95,dunsby96b,LC-sw,LC-scalcmb,gebbie98,gebbie98b}
field~\cite{stoeger95b}--\cite{gebbie98b},
we have found the 1+3 covariant formulation of
hydrodynamics~\cite{ehlers93,ellis71} and radiative transport
theory~\cite{ellis83,thorne81} to be well suited to this task.

In this contribution we begin by reviewing the exact 1+3 covariant theory of
polarized radiative transport in the presence of Thomson scattering (the
dominant scattering mechanism for CMB photons over the epoch of interest). This
work builds on the covariant approach to kinetic theory developed
in~\cite{ellis83,thorne81}. The angular dependence of the intensity and
polarization are conveniently described
in terms of projected symmetric trace free (PSTF) tensors. The multipole
equations governing the dynamics of these radiation variables are given in
linearised form around an almost-FRW universe, and generic features of their
solution are discussed qualitatively. By further decomposing the linearised
multipole equations in harmonic functions, we arrive at a complete set of
gauge-invariant perturbation equations in a form which
is convenient for numerical implementation, but which still preserves the
physical transparency of the 1+3 covariant approach. We present the
mode-expanded perturbation equations for scalar (density clumping) modes, and
discuss briefly their numerical implementation and some numerical results.

We employ standard general relativity and adopt a $(+---)$ signature for the
metric $g_{ab}$. Our conventions for the Riemann and Ricci tensors are fixed by
$[\nabla_a,\nabla_b]u_c={R_{abc}}^d u_d$ and $R_{ac}={R_{abc}}^b$.
The spacetime alternating tensor is denoted $\eta_{abcd}$.
Round brackets denote symmetrisation on the enclosed indices, square brackets
denote antisymmetrisation, and angle brackets denote the PSTF part
(the projection is with $h_{ab}=g_{ab}-u_a u_b$, where $u_a$ is
the fundamental velocity of the 1+3 approach). Overdots denote the action of
$u^a \nabla_a$, and the totally projected covariant ``derivative'' is denoted
by $D_a$.

\section{Covariant Radiative Transport Theory}

In the 1+3 covariant approach to cosmology (see~\cite{ellis98} for a
recent review), the cosmological model is described in terms of geometric
variables that have a direct interpretation in terms of the observations
of a preferred set of observers who are comoving with the fundamental velocity
field $u^a$. We shall refer to the choice of $u^a$ as a frame
choice\footnote{The issue of how to specify $u^a$ in an appropriate
physical manner is discussed in~\cite{gebbie98b,ellis98}.}. We
consider a diffuse radiation field (the CMB) observed from the $u^a$
frame. The Stokes parameters of the radiation in some solid angle $d\Omega$
and some photon energy range $dE$, along the
unit projected vector $e^a$, can be combined into a second-rank polarization
tensor $P_{ab}(E,e)$ which is projected relative to $u^a$ and $e^a$ on both
indices. The non-vanishing components of $P_{ab}(E,e)$ on an orthonormal
tetrad with the 0-direction aligned with $u^a$ and the 3-direction with $e^a$
are defined in terms of the Stokes parameters via:
\begin{align}
P_{11} & = {\tfrac{1}{2}} [I(E,e) + Q(E,e)] \notag \\
P_{12} & = {\tfrac{1}{2}} [U(E,e) + V(E,e)] \notag \\
P_{21} & = {\tfrac{1}{2}} [U(E,e) - V(E,e)] \notag \\
P_{22} & = {\tfrac{1}{2}} [I(E,e) - Q(E,e)].
\end{align}
Our notation for the Stokes parameters follows~\cite{kosowsky96}.
The transformation properties of the Stokes parameters under rotation of the
1- and 2-directions ensures that $P_{ab}(E,e)$, as defined, is tensor-valued.
Introducing a screen projection tensor $\clh_{ab}$,
\begin{equation}
\clh_{ab} \equiv h_{ab} + e_a e_b,
\end{equation}
the polarization tensor can be written in irreducible form
\begin{equation}
P_{ab}(E,e) = - {\tfrac{1}{2}} I(E,e) \clh_{ab}
+ \clp_{ab}(E,e) - {\tfrac{1}{2}} V(E,e) \epsilon_{abc} e^c,
\label{irred}
\end{equation}
where $\epsilon_{abc} \equiv \eta_{abcd} u^d$ and $\clp_{ab}(E,e)$ is
a PSTF tensor which is orthogonal to $e^a$:
\begin{equation}
\clp_{ab}(E,e) = \clh_a^c \clp_{cb}(E,e).
\end{equation}

\subsection{The Multipole Decomposition}

For many applications it is convenient to describe the angular dependence
of the polarization tensor in terms of PSTF tensor-valued multipole
moments. Since $I(E,e)$ and $V(E,e)$ are scalar fields on the sphere they
can be expanded as follows:
\begin{align}
I(E,e) & = \sum_{l=0}^{\infty} I_{A_l}(E) e^{A_l} \\
V(E,e) & = \sum_{l=0}^{\infty} V_{A_l}(E) e^{A_l},
\end{align}
where $I_{A_l}(E)$ and $V_{A_l}(E)$ are energy-dependent PSTF tensors,
and we have introduced the index notation $A_l \equiv a_1 \dots a_l$
with $e^{A_l} \equiv e^{a_1} \dots e^{a_l}$. The expansions of
$I(E,e)$ and $V(E,e)$ can be inverted to determine the multipoles, for example
\begin{equation}
I_{A_l}(E) = \frac{1}{\Delta_l} \int d\Omega\, I(E,e) e_{\langle A_l \rangle},
\quad \mbox{with} \quad \Delta_l \equiv \frac{4\pi (-2)^l (l!)^2}{(2l+1)!}.
\end{equation}

For the linear polarization tensor $\clp_{ab}(E,e)$ we must expand in tensor
spherical harmonics that are projected perpendicular to
$e^a$ (tensors projected perpendicular to $e^a$, such as $\clp_{ab}(E,e)$,
are denoted transverse in~\cite{thorne80}). It is most convenient to use
pure-spin PSTF harmonics~\cite{thorne80}, so that
\begin{equation}
\clp_{a_1 a_2}(E,e) = \sum_{l=2}^{\infty} [\cle_{a_1 a_2 B_{l-2}}
e^{B_{l-2}}]^{TT}
+ \sum_{l=2}^{\infty} [e_{c_1} {\epsilon^{c_1 c_2}}_{(a_1}
\clb_{a_2) c_2 B_{l-2}} e^{B_{l-2}}]^{TT}.
\label{eq_1}
\end{equation}
Coordinate-dependent versions of this expansion were first
discussed in the context of all-sky CMB polarization data
in~\cite{kamion97,seljak97}.
In equation~\eqref{eq_1} we have left the energy dependence of the PSTF
tensors $\cle_{A_l}(E)$ and $\clb_{A_l}(E)$ implicit. The superscript
$\scriptstyle{TT}$ denotes the transverse traceless part:
\begin{equation}
[A_{ab}]^{TT} \equiv \clh_a^{c_1} \clh_b^{c_2} A_{c_1 c_2}
- {\tfrac{1}{2}} \clh_{ab} \clh^{c_1 c_2} A_{c_1 c_2},
\end{equation}
where $A_{ab}$ is a second-rank tensor. The two summations in
equation~\eqref{eq_1} define the electric and magnetic parts of
$\clp_{ab}(E,e)$. The $l$-th term in the electric part has parity
$(-1)^l$, while the $l$-th term in the magnetic part has parity
$(-1)^{l-1}$. The electric and magnetic multipoles can be determined
from the polarization tensor with the following inversion formulae:
\begin{align}
\cle_{A_l}(E) & = \frac{1}{\Delta_l} \frac{2l (l-1)}{(l+1)(l+2)}
\int d\Omega \, e_{\langle A_{l-2}} \clp_{a_l a_{l-1}\rangle } (E,e),\\
- \clb_{A_l}(E) & = \frac{1}{\Delta_l} \frac{2l (l-1)}{(l+1)(l+2)}
\int d\Omega \, e_b {\epsilon^{bc}}_{\langle a_l} e_{A_{l-2}}
\clp_{a_{l-1} \rangle c} (E,e),
\end{align}
which follow from the orthogonality of the $e_{\langle A_l \rangle}$.
We shall also make use of the energy-integrated multipoles,
$I_{A_l}$, $\cle_{A_l}$, $\clb_{A_l}$ and $V_{A_l}$, where, for example,
\begin{equation}
I_{A_l} \equiv \Delta_l \int dE\, I_{A_l}(E).
\end{equation}
It follows that the three lowest energy-integrated multipoles of $I(E,e)$
give the radiation energy density, momentum density and anisotropic stress
respectively:
\begin{equation}
I = \rho^{(\gamma)}, \qquad I_a = q^{(\gamma)}_a, \qquad I_{ab}=
\pi^{(\gamma)}_{ab}.
\end{equation}

The scale dependence of intensity and polarization correlations on the CMB sky
are described by their angular power spectra. In a statistically isotropic
ensemble of CMB skies, the power spectrum of intensity
anisotropies $C^{II}_l$ is defined by the ensemble average~\cite{gebbie98}
\begin{equation}
\left(\frac{\pi}{I}\right)^2 \langle I_{A_l} I^{B_{l'}} \rangle
= \Delta_l C^{II}_l\delta_{ll'} h_{\langle A_l \rangle}^{\langle B_{l}
\rangle},
\end{equation}
where $h_{\langle A_l \rangle}^{\langle B_{l}\rangle}
\equiv h_{\langle a_1}^{\langle b_1} \dots  h_{a_l \rangle}^{b_l \rangle}$.
In the limit of almost-isotropic radiation, the $I_{A_l}$ are related to the
gauge-invariant bolometric temperature perturbation (from the all-sky mean)
$\delta_T(e)$ via~\cite{maartens95}
\begin{equation}
\delta_T(e)= \frac{\pi}{I} \sum_{l=1}^{\infty} \frac{1}{\Delta_l}
I_{A_l} e^{A_l},
\end{equation}
from which we recover the usual Legendre expansion of the temperature
correlation function:
\begin{equation}
\langle \delta_T(e) \delta_T(e')\rangle = \sum_{l=1}^{\infty}
\frac{(2l+1)}{4\pi} C_l^{II} P_l(-e^a e'_a).
\end{equation}

Power spectra for the polarization and the cross-correlation between
electric polarization and intensity anisotropies can be defined in
a similar manner to $C_l^{II}$. However, for consistency with previous
approaches based on an expansion in
spin-weighted harmonics~\cite{seljak97}, with the normalisation conventions
adopted here it is necessary to include an additional factor
of $\surd[l(l-1)/(l+1)(l+2)]$ in the definition of $C_l$ for each factor
of the polarization. For example,
\begin{equation}
\left(\frac{\pi}{I}\right)^2 \langle \cle_{A_l}\cle^{B_{l'}}
\rangle = \frac{l(l-1)}{(l+2)(l+1)}\Delta_l C_l^{\cle\cle} \delta_{ll'}
h_{\langle A_l \rangle}^{\langle B_{l} \rangle}.
\end{equation}
Note there cannot be any correlation between $\clb_{A_l}$ and
either of $\cle_{A_l}$ or $I_{A_l}$ in a parity symmetric ensemble.
As noted in~\cite{seljak97}, the definitions of the polarization power spectra
adopted here are independent of the choice of polarization basis vectors, which
has significant advantages over correlating the Stokes parameters directly.

\subsection{The Boltzmann Equation}

The evolution of the polarization tensor $P_{ab}(E,e)$ along the path
of the radiation in phase space is described by the (collisional)
Boltzmann equation
\begin{equation}
\cll[ E^{-3} P_{ab}(E,e)] = K_{ab}(E,e).
\label{boltz1}
\end{equation}
The appropriate Liouville operator $\cll$ for the transverse tensor
$P_{ab}(E,e)$ acts according to
\begin{equation}
\cll P_{ab}(E,e) = \clh_a^{c_1} \clh_b^{c_2} p^d \nabla_d P_{c_1 c_2}(E,e),
\end{equation}
where the derivative is along the photon path $\{x^a(\lambda),p^a(\lambda)\}$
in phase space. Here, $p^a = E(u^a + e^a)$ is the photon momentum and
$\lambda$ is an affine parameter with $p^a = dx^a / d\lambda$.
The scattering tensor $K_{ab}(E,e)$ on the right-hand side of
equation~\eqref{boltz1} is transverse. Its exact form for Thomson scattering
is given in Section~\ref{sec_thom}.

The irreducible decomposition of $P_{ab}(E,e)$ into $I(E,e)$,
$V(E,e)$ and $\clp_{ab}(E,e)$, given as equation~\eqref{irred}, is preserved
by the action of the Liouville operator, so that
\begin{multline}
\cll[ E^{-3} P_{ab}(E,e)] = -\frac{1}{2}  \frac{d}{d\lambda}
[E^{-3} I(E,e)] \clh_{ab} + \cll[ E^{-3} \clp_{ab}(E,e)] \\ - \frac{1}{2}
\frac{d}{d\lambda}[E^{-3} V(E,e)]\epsilon_{abc} e^c.
\end{multline}
In the absence of scattering ($K_{ab}(E,e)=0$), we recover the well-known
results that the occupation number $I(E,e)/E^3$, degree of
circular polarization $V(E,e)/I(E,e)$, and degree of linear
polarization $\surd[2 \clp_{ab}(E,e) \clp^{ab}(E,e)]/I(E,e)$ are independently
conserved along the path of the radiation.

\subsubsection{Transformation Laws Under Changes of Frame}

The 1+3 covariant approach considers observations made from the viewpoint
of a set of observers moving with the fundamental velocity $u^a$. Often, it is
useful to have available the transformation laws relating these observations
to those made with respect to a different velocity field $\tilde{u}^a=
\gamma (u^a + v^a)$. Here, $v^a$ is the projected relative velocity (in the
$u^a$ frame), and $\gamma$ is the associated Lorentz factor.

For a given photon momentum $p^a=E(u^a + e^a)$, the energy $\tilde{E}$ and
direction of propagation $\tilde{e}^a$ in the $\tilde{u}^a$ frame are
given by
\begin{align}
\tilde{E} & = \gamma E (1+ e^a v_a) \\
\tilde{e}^a & = [\gamma(1+e^b v_b)]^{-1}(e^a + u^a) - \gamma (u^a + v^a).
\end{align}
The screen projection tensor $\tilde{\clh}_{ab}$ in the $\tilde{u}^a$ frame
is related to the equivalent tensor in the $u^a$ frame by
\begin{equation}
\tilde{\clh}_{ab} = \clh_{ab} - \frac{2\gamma}{\tilde{E}}p_{(a}\clh_{b)c}
v^c + \left(\frac{\gamma}{\tilde{E}}\right)^2 p_a p_b \clh^{c_1 c_2} v_{c_1}
v_{c_2}.
\end{equation}
The polarization tensor in the $\tilde{u}^a$ frame,
$\tilde{P}_{ab}(\tilde{E},\tilde{e})$, is related to $P_{ab}(E,e)$ by
\begin{equation}
\tilde{E}^{-3}\tilde{P}_{ab}(\tilde{E},\tilde{e})
= E^{-3} \tilde{\clh}_a^{c_1} \tilde{\clh}_b^{c_2}
P_{c_1 c_2}(E,e).
\end{equation}
From this transformation law we recover the frame-invariance of
$I(E,e)/E^3$, and the degrees of circular and linear polarization.

It is straightforward to verify that under changes of frame the action of the
Liouville operator transforms according to
\begin{equation}
\tilde{\cll} [\tilde{E}^{-3}\tilde{P}_{ab}(\tilde{E},\tilde{e})]
=  \tilde{\clh}_a^{c_1} \tilde{\clh}_b^{c_2}
\cll [E^{-3} P_{c_1 c_2}(E,e)],
\end{equation}
so that the collision tensor in the Boltzmann equation must transform as
\begin{equation}
\tilde{K}_{ab}(\tilde{E},\tilde{e}) = \tilde{\clh}_a^{c_1} \tilde{\clh}_b^{c_2}
K_{c_1 c_2}(E,e).
\end{equation}
This transformation law provides a useful means of generating the
scattering tensor for a general choice of frame given its form in some
particular frame. (It is often the case that the scattering process picks out
a frame in which $K_{ab}(E,e)$ is particularly simple.)
The transformation laws for the radiation multipoles $I_{A_l}(E)$,
$V_{A_l}(E)$, $\cle_{A_l}(E)$ and $\clb_{A_l}(E)$ are non-local in energy
and show multipole coupling due to relativistic beaming. Since the detailed
form of the multipole transformation laws is not needed for the
calculation of linearised CMB anisotropies and polarization, we shall not give
them here.

\subsubsection{$K_{ab}(E,e)$ for Thomson Scattering}
\label{sec_thom}

Over the epoch of interest for the generation of CMB anisotropies and
polarization, the dominant scattering mechanism for CMB photons is Compton
scattering off free electrons. Well after electron/positron annihilation
it is a good approximation to ignore electron recoil and the small thermal
effects due to the electron distribution function. In this limit it is
convenient to work initially in the rest frame of the tightly-coupled
electron/baryon plasma, the four-velocity of which we denote by $\tilde{u}^a$.

The scattering tensor $K_{ab}(E,e)$ can be deduced from the
standard result for the Thomson scattering kernel
expressed in terms of the Stokes parameters
(see, for example,~\cite{chand_rad}). Assuming an unpolarized
distribution of electrons, we find the following expression for the
scattering tensor in the $\tilde{u}^a$ frame in 1+3 covariant irreducible
form:
\begin{align}
\tilde{E}^2 \tilde{K}_{ab}(\tilde{E},\tilde{e}) &= n_e \sigt
\biggl\{-{\frac{1}{2}} \tilde{\clh}_{ab} \Bigl[-\tilde{I}(\tilde{E},\tilde{e})
+ \tilde{I}(\tilde{E}) \notag \\
&\phantom{xx} + {\frac{1}{10}} \tilde{I}_{c_1 c_2}(\tilde{E})\tilde{e}^{c_1}
\tilde{e}^{c_2} + {\frac{3}{5}} \tilde{\cle}_{c_1 c_2}(\tilde{E})
\tilde{e}^{c_1} \tilde{e}^{c_2}\Bigr]\notag \\
&\phantom{xx} + \Bigl[- \tilde{\clp}_{ab}(\tilde{E},\tilde{e}) +
{\frac{1}{10}} [\tilde{I}_{ab}(\tilde{E})]^{TT} + {\frac{3}{5}}
[\tilde{\cle}_{ab}(\tilde{E})]^{TT} \Bigr] \notag \\
&\phantom{xx} - {\frac{1}{2}} \tilde{\epsilon}_{abc} \tilde{e}^c
\Bigl[- \tilde{V}(\tilde{E},\tilde{e}) + {\frac{1}{2}}
\tilde{V}_d(\tilde{E}) \tilde{e}^d\Bigr]\biggr\},
\label{thomxc}
\end{align}
where $n_e$ is the electron number density in the $\tilde{u}^a$ frame,
and $\sigma_T$ is the Thomson cross section.
Equation~\eqref{thomxc} is an exact expression for the scattering tensor in
the rest frame of the electrons. The first set of terms in square brackets
source the intensity evolution, the second set source the evolution of the
linear polarization, while the final set source the circular polarization.
Note that the source terms for the intensity have no monopole ($l=0$) moment
in the $\tilde{u}^a$ frame, since there is no energy transfer in the
Thomson limit in the electron rest frame. Note also how the quadrupole
($l=2$) moment of the electric polarization sources the intensity, and the
quadrupole moment of the intensity sources the electric polarization.
(It is for this reason that polarization should be included in accurate
CMB codes even if only the temperature information is of interest.)
Equation~\eqref{thomxc} is equivalent to the scattering terms derived recently
within the context of the ``total angular momentum'' advocated by Hu and
collaborators~\cite{hu97a,hu98}. However, unlike the results
in~\cite{hu97a,hu98},
equation~\eqref{thomxc} is free from any (spatial) harmonic decomposition
of the radiation variables.

Transforming the scattering term to an arbitrary frame, and integrating over
energies, we find the exact evolution of the intensity:
\begin{align}
\int dE\, E^2 \cll[E^{-3} I(E,e)] &=
- n_e \sigt \gamma (1+ e^a v_a) \int dE\, I(E,e) \notag \\
&\phantom{xx}+ {\frac{1}{4\pi}} n_e \sigt [\gamma(1+e^a v_a)]^{-3} (
\tilde{I} +\tilde{\zeta}_{c_1 c_2} 
\tilde{e}^{c_1} \tilde{e}^{c_2}),
\end{align}
the linear polarization:
\begin{align}
\int dE\, E^2 \cll[E^{-3} \clp_{ab}(E,e)] &=
- n_e \sigt \gamma (1+ e^c v_c) \int dE\, \clp_{ab}(E,e) \notag \\
&\phantom{xx}+ {\frac{1}{4\pi}} n_e \sigt [\gamma(1+e^c v_c)]^{-3}
[\tilde{\clh}^{c_1}_a \tilde{\clh}^{c_2}_b
\tilde{\zeta}_{c_1 c_2}]^{TT},
\end{align}
and the circular polarization:
\begin{align}
\int dE\, E^2 \cll[E^{-3} V(E,e)] &=
- n_e \sigt \gamma (1+ e^a v_a) \int dE\, V(E,e) \notag \\
&\phantom{xx}- {\frac{3}{8\pi}}n_e \sigt [\gamma(1+e^a v_a)]^{-3}
\tilde{V}_b \tilde{e}^b.
\label{thomcirc}
\end{align}
Here, tildes refer to quantities in the rest frame of the electrons, and
\begin{equation}
\tilde{\zeta}_{ab} \equiv  {\frac{3}{4}} \tilde{I}_{ab}
+ {\frac{9}{2}} \tilde{\cle}_{ab}.
\end{equation}
Expressions for $\tilde{I}$, $\tilde{I}_{ab}$, and $\tilde{\cle}_{ab}$
in terms of quantities in the $u^a$ frame can be derived from the
multipole transformation laws if required.

Equation~\eqref{thomcirc} demonstrates that circular polarization is not
sourced by Thomson scattering off unpolarized electrons, so that in the
absence of polarising agents, such as primordial magnetic fields,
we should not expect the primordial component of
the microwave sky to have any circular polarization. In light of these
remarks, we shall not consider the $V$ Stokes parameter any further here.
(Some effects of primordial magnetic fields on the CMB are discussed
in~\cite{loeb96,durrer98}.)

\section{Linearisation Around a FRW Model}

The observed isotropy of the CMB (the anisotropies are at the level
of only one part in $10^5$~\cite{smoot92}) combines with the Copernican
assumption to imply that on large scales the universe is well approximated by
perturbing a FRW model~\cite{stoeger95b,maartens95}.
In the 1+3 covariant and gauge-invariant approach
to studying perturbations in cosmology~\cite{haw66,ellis89a,ellis89b},
the perturbed model is investigated in terms of covariant variables
which are defined in the real universe (as opposed to the background model),
and which vanish identically in the background model. With a physical
choice for the fundamental velocity $u^a$, the approach is gauge-invariant
to all orders. Given a smallness parameter $\epsilon$ for the almost-FRW model,
the covariant variables that describe inhomogeneity and anisotropy are at
most $O(\epsilon)$. Important examples of $O(\epsilon)$ variables in a generic
almost-FRW spacetime include the kinematic variables derived from $u^a$:
\begin{align}
& \text{shear} & \sigma_{ab} &= D_{\langle a}u_{b \rangle} \notag \\
& \text{vorticity} & \omega_{ab} &= D_{[a}u_{b]} \notag \\
& \text{acceleration} & A_a &= \dot{u}_a \notag \\
& \text{expansion inhomogeneity} & \clz_a &= S D_a \Theta \, ;
\end{align}
the matter variables derived from the matter stress-energy tensor $T_{ab}$:
\begin{align}
& \text{momentum density} & q_a &= T_{\langle a \rangle b} u^b \notag \\
& \text{anisotropic stress} & \pi_{ab} &= T_{\langle a b\rangle} \notag \\
& \text{density inhomogeneity} & \clx_a &= S D_a \rho /\rho \, ;
\end{align}
the Weyl tensor $C_{abcd}$ and hence its electric and magnetic parts:
\begin{align}
& \text{electric} & E_{ab} &= C_{acbd} u^c u^d \notag \\
& \text{magnetic} & H_{ab} &= -{\tfrac{1}{2}}\epsilon_{acd}
{C_{be}}^{cd} u^e \, .
\end{align}
Here, $\Theta = \nabla_a u^a$ is the fractional volume expansion rate,
$\rho = T_{ab} u^a u^b$ is the matter energy density, and $S$ is a covariantly
defined scale factor, $\dot{S}/S = \Theta /3$ with $D_a S = O(\epsilon)$.
The linearised perturbation equations for the $O(\epsilon)$ variables are
derived from the exact 1+3 covariant hydrodynamic and gravito-electromagnetic
equations (see, for example,~\cite{ellis98}).
The equation set closes when supplemented by constitutive relations for the
matter variables.

\subsection{Linearised Multipole Equations}
\label{multipole}

The equations of radiative transfer provide the constitutive relations
for the radiation sector of the cosmological model. It turns out to be most
convenient to work directly with the evolution equations for the radiation
angular multipoles. The exact multipole equations can be derived from
equation~\eqref{boltz1}. The equations for the intensity multipoles
$I_{A_l}(E)$ are derived in~\cite{gebbie98b,ellis83,thorne81};
those for the polarization multipoles
are given in~\cite{chall99b}. Here, we shall only make use of the linearised
multipole equations. Since Thomson scattering does not produce any spectral
distortion at linear order (this follows from equation~\eqref{thomxc}), it
is sufficient to consider only the energy-integrated multipoles
$I_{A_l}$, $\cle_{A_l}$ and $\clb_{A_l}$. Noting that $\cle_{A_l}$ and
$\clb_{A_l}$ are $O(\epsilon)$ for all $l\geq 2$ in an almost-FRW universe,
while $I_{A_l}$ are $O(\epsilon)$ for $l>0$, we find the following linearised
evolution equations for the electric polarization\footnote{We use the
convention that $\cle_{A_l}$ and $\clb_{A_l}$ vanish for $l<2$.}:
\begin{align}
\dot{\cle}_{A_l} + \frac{4}{3}\Theta \cle_{A_l} &+
\frac{(l+3)(l-1)}{(l+1)^2} D^b \cle_{b A_l} - \frac{l}{(2l+1)}
D_{\langle a_l} \cle_{A_{l-1} \rangle}\notag
\\ & - \frac{2}{(l+1)}
\curl \clb_{A_l} = -n_e \sigt \left(\cle_{A_l} - \frac{2}{15}\delta_{l2}
\zeta_{ab} \right),
\label{elecmtp}
\end{align}
and for the magnetic polarization:
\begin{align}
\dot{\clb}_{A_l} + \frac{4}{3}\Theta \clb_{A_l} &+
\frac{(l+3)(l-1)}{(l+1)^2} D^b \clb_{b A_l} - \frac{l}{(2l+1)}
D_{\langle a_l} \clb_{A_{l-1} \rangle}\notag
\\ & + \frac{2}{(l+1)} \curl \cle_{A_l} = -n_e \sigt \clb_{A_l}.
\label{magmtp}
\end{align}
Here, the curl of a rank-$l$ tensor $S_{A_l}$ is defined by
\begin{equation}
\curl S_{A_l} \equiv {\epsilon_{c_1 c_2 (a_l}} D^{c_1} {S_{A_{l-1})}}^{c_2},
\end{equation}
which is PSTF for $S_{A_l}$ PSTF. Equations~\eqref{elecmtp} and~\eqref{magmtp}
reveal that $\cle_{A_l}$ and $\clb_{A_l}$ are coupled through curl terms
(as with Maxwell's equations), and that only the electric quadrupole
$\cle_{ab}$ has inhomogeneous scattering terms in its evolution equation.
These observations are crucial for the prospect of detecting the imprint
of primordial gravitational waves on the CMB (see Section~\ref{sec_scalar}).

The linearised intensity multipole equations are summarised by
\begin{multline}
\dot{I}_{A_l} + {\frac{4}{3}} \Theta I_{A_l} + D^b I_{b A_l}
- {\frac{l}{(2l+1)}} D_{\langle a_l} I_{A_{l-1} \rangle}
+ {\frac{4}{3}}I A_{a_1} \delta_{l1} - {\frac{8}{15}}I \sigma_{a_1 a_2}
\delta_{l2} \\
= -n_e \sigt \left(I_{A_l} - {\frac{4}{3}}I v_{a_1} \delta_{l1}
- {\frac{2}{15}}I \zeta_{a_1 a_2} \delta_{l2} \right), \quad l\geq 1,
\label{inmtp}
\end{multline}
where $v^a$ is the projected relative velocity of the electron rest fame.
The intensity monopole (radiation energy density) evolves according to
\begin{equation}
\dot{I} + {\frac{4}{3}} \Theta I + D^a I_a = 0,
\end{equation}
which implies the following evolution equation for the comoving projected
gradient of the energy density:
\begin{equation}
(SD_a I)^{\cdot} + {\frac{4}{3}}\Theta S D_a I + S D_a D^b I_b +
{\frac{4}{3}} I \clz_a - {\frac{4}{3}}\Theta S I A_a = 0.
\label{denmtp}
\end{equation}

The linearised multipole
equations~\eqref{elecmtp},~\eqref{magmtp},~\eqref{inmtp} and~\eqref{denmtp}
provide a complete description of linearised CMB physics in an almost-FRW
universe. From these equations one can infer the model-independent behaviour
of the anisotropy and polarization. Prior to recombination, the photon
mean free time is short compared to the characteristic expansion time of the
universe. On scales much larger than the photon mean free path, the photons
are tightly-coupled to the electron/baryon plasma, and the combined
system behaves like a fluid. On super-horizon scales $D_a I$ grows due to
gravitational instability, while on sub-horizon scales power oscillates
between $D_a I$ and $I_a$ (acoustic oscillations). The fluid approximation
breaks down inside the diffusion scale~\cite{silk67}, where higher-$l$
intensity multipoles are excited at the expense of damping the acoustic
oscillations. During recombination the mean free path of the photons rises
to effectively infinity, with the result that
power is transported to higher-$l$ intensity multipoles through the
free-streaming projection of $D_a I$, the electron peculiar velocity
$v_a$, and the combination of intensity and electric polarization quadrupoles
$\zeta_{ab}$ from the last scattering surface. Power is also transferred to the
intensity multipoles through variations of redshift since last scattering along
different lines of sight. The electric polarization quadrupole $\cle_{ab}$
is generated through recombination from the growing intensity quadrupole
anisotropy. This mechanism ceases at last scattering, after which the
power in the electric quadrupole is redistributed amongst the higher-$l$
electric and magnetic multipoles via free-streaming. Any reionisation tends to
damp the primary intensity anisotropies while boosting the polarization signal,
as well as generating new secondary intensity anisotropies
(see~\cite{haiman99} for a recent review).

\subsection{Scalar Perturbations}
\label{sec_scalar}

Up to this point our discussion has been concerned with arbitrary (small)
perturbations around a FRW model. In particular, unlike most other
perturbative approaches (for example,~\cite{bardeen80}), we have not
had to split the perturbations into scalar, vector or tensor modes
(see, for example,~\cite{lifshitz46}) to derive the basic, gauge-invariant
perturbation equations.
However, for detailed calculation it is convenient to make a
non-local mode decomposition of the $O(\epsilon)$ covariant variables
to decouple the temporal and spatial dependencies of the 1+3 equations.
The decomposition can be performed in linear theory in a fully covariant
manner, which leads to a physically transparent, gauge-invariant perturbation
theory. Here we shall only give a brief review of the procedure for scalar
modes (those which describe density clumping)~\cite{LC-scalcmb,ellis89a}.
Vector and tensor perturbations are treated in a similar manner.

By definition, in scalar modes the $O(\epsilon)$ variables are constructed
from $O(\epsilon)$ scalar potentials. It follows from the constraint
equations of 1+3 covariant hydrodynamics~\cite{ellis98} that the vorticity
vanishes at $O(\epsilon)$ in a scalar mode. Furthermore, since all curls vanish
by construction, the constraint equations of 1+3
gravito-electromagnetism~\cite{ellis98,maartens98} constrain the magnetic part
of the Weyl tensor to vanish at linear order. The fact that $\curl \cle_{A_l}$
and $\curl \clb_{A_l}$ vanish at linear order removes the coupling between
the electric and magnetic polarization multipoles in their
evolution equations~\eqref{elecmtp} and~\eqref{magmtp}. It follows that the
hierarchy for the $\clb_{A_l}$ has
no inhomogeneous terms present, so that scalar modes do not generate
magnetic polarization~\cite{kamion97,seljak97}. Any detection of
primordial magnetic polarization would provide evidence of vector or tensor
modes. This proves to be a much more promising approach than relying on their
cosmic variance limited temperature signature~\cite{kamion98}.

The scalar potentials from which the $O(\epsilon)$ covariant variables are
derived may be expanded in the complete set of scalar eigenfunctions
$Q^{(k)}$ of the comoving Laplacian:
\begin{equation}
D_a D^a Q^{(k)} = {\frac{k^2}{S^2}} Q^{(k)},
\end{equation}
with $\dot{Q}^{(k)} = O(\epsilon)$. Rank-$l$ PSTF tensors, which are
covariantly constant along the integral curves of $u^a$,
can be constructed from the $Q^{(k)}$:
\begin{equation}
Q^{(k)}_{A_l} \equiv \left({\frac{S}{k}}\right)^{l}
D_{\langle a_1} \dots D_{a_l \rangle} Q^{(k)}.
\end{equation}
The $Q^{(k)}_{A_l}$ can be used to expand the radiation multipoles,
so that, symbolically,
\begin{equation}
I_{A_l} = I\sum_k I_k^{(l)} Q^{(k)}_{A_l}\, , \quad
\cle_{A_l} = I \sum_k \cle_k^{(l)} Q^{(k)}_{A_l},
\end{equation}
where the $I^{(l)}_k$ and $\cle^{(l)}_k$ are $O(\epsilon)$ scalars with
$O(\epsilon^2)$ projected gradients. It proves convenient to use the notation
$I^{(0)}_k$ for the scalar coefficients in the expansion of $D_a I$:
$D_a I = I \sum_k (k/S) I^{(0)}_k Q^{(k)}_a$, so that the mode-expanded
intensity multipole equations take the form
\begin{multline}
\dot{I}^{(l)}_k - \frac{k}{S} \biggr\{
\frac{l}{(2l+1)} I_k^{(l-1)} - \frac{(l+1)}{(2l+1)}
\left[1-l(l+2)\frac{K}{k^2}\right]I_k^{(l+1)}\biggr\} \\
+ {\frac{4}{3}} {\frac{k}{S}} A_k \delta_{l1} -
{\frac{8}{15}} {\frac{k}{S}} \sigma_k \delta_{l2} =
-n_e \sigt \left(I^{(l)}_k - {\frac{4}{3}} v_k \delta_{l1} -
{\frac{2}{15}} \zeta_k \delta_{l2} \right).
\end{multline}
Here $A_k$, $\sigma_k$, $v_k$ and $\zeta_k$ are the dimensionless
coefficients in the mode expansions of $A_a$, $\sigma_{ab}$, $v_a$ and
$\zeta_{ab}$, and $6K/S^2$ is the zero-order 3-curvature scalar.
The evolution equation for $I^{(0)}_k$ follows from
equation~\eqref{denmtp}:
\begin{equation}
\dot{I}^{(0)}_k + {\frac{k}{S}}I^{(1)}_k + {\frac{4}{3}}{\frac{k}{S}}
\clz_k - {\frac{4}{3}}\Theta A_k = 0,
\end{equation}
where $\clz_k$ are the dimensionless coefficients in the expansion of
$\clz_a$. For the electric polarization we find
\begin{multline}
\dot{\cle}^{(l)}_k - \frac{k}{S} \biggr\{
\frac{l}{(2l+1)} \cle_k^{(l-1)} - \frac{(l+3)(l-1)}{(2l+1)(l+1)}
\left[1-l(l+2)\frac{K}{k^2}\right]\cle_k^{(l+1)}\biggr\} \\
= - n_e \sigt\left(\cle^{(l)}_k - \frac{2}{15}\zeta_k \delta_{l2}\right).
\end{multline}

At this point we have essentially recovered the well-known multipole
equations for scalar perturbations in an almost-FRW
model~\cite{hu98,seljak98}. Arguably, the equations given here are somewhat
more transparent than their gauge-dependent counterparts in alternative
approaches. For example, the intensity monopole $I^{(0)}_k$ describes the
spatial inhomogeneity of the radiation energy density, which is of clear
physical significance, rather than the gauge-dependent difference between
the density in the real universe and the background model. The
mode-expanded multipole equations can be integrated directly to
determine the anisotropy and polarization for some given initial conditions.
To close the equations one must also integrate the mode-expanded
hydrodynamical equations for the kinematic variables that
source the intensity equations through redshift effects~\cite{LC-scalcmb}.
Since much of the complexity of the coupled multipole equations arises
because they describe evolution along the timelike integral curves of
$u^a$, rather than along the lightlike trajectory of the radiation,
it is more efficient to employ integral solutions to the multipole
equations rather than integrate a large number of them
directly~\cite{seljak96}.
The integral solutions consist of simple source functions integrated against
known geometric projection functions. The integral solutions for
scalar and tensor modes are given in~\cite{chall99b}; see also~\cite{hu98}.

\begin{figure}[t!]
\begin{center}
\epsfig{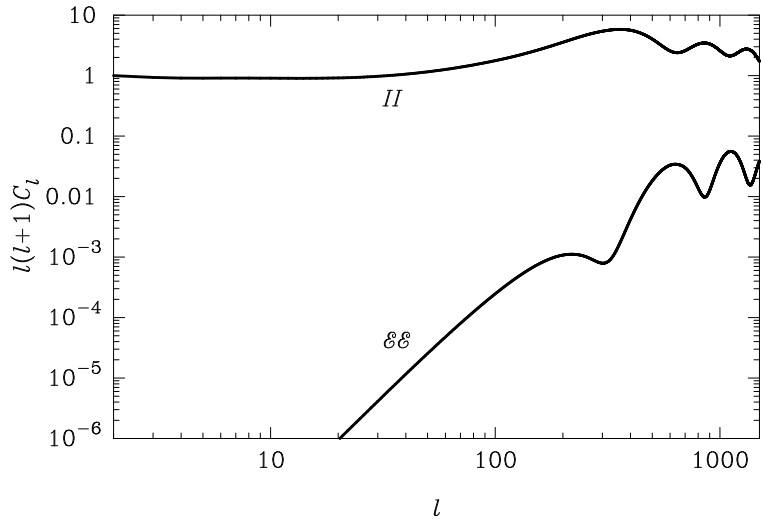}
\end{center}
\caption{Intensity $C_l^{II}$ and polarization $C_l^{\cle\cle}$
power spectra from scalar modes in a scale-invariant, adiabatic, open cold
dark matter (CDM) model. The baryon fraction is $\Omega_b=0.05$, the CDM
fraction is $\Omega_c=0.4$, the Hubble constant is $H_0=50\kmsmpc$,
and there is no cosmological constant nor reionisation.} 
\label{fig1}
\end{figure}

An example of the intensity and polarization power spectra from scalar
perturbations in an open model is given in Figure~\ref{fig1}. The code to
produce the figure was based on CMBFAST~\cite{seljak96};
the necessary modifications to solve the covariant equations were implemented
by Antony Lewis. 

\section{Conclusion}

The 1+3 covariant formulation of radiative transfer provides a
convenient, exact framework within which to study CMB physics. Although we
have only considered the linearised form of the transfer equations in our
discussion of the anisotropy and polarization in an almost-FRW universe,
the formalism is of wider applicability. For example, the approach
appears well suited to a systematic study of second-order geometric and
scattering effects~\cite{gebbie98b}. In the application to linearised
CMB physics in an almost-FRW universe, the 1+3 covariant perturbative
approach shares many of the advantages of the ``total angular momentum''
method~\cite{hu97a,hu98}. These include physical transparency,
the explicitly quadrupole nature of the
radiative source terms in the Boltzmann equation that arise from the
angular and polarization dependence of the Thomson cross section, and the
direct manner in which the angular multipoles describe the angular scale
of anisotropy on the sky. However, the 1+3 covariant method goes further by
ensuring manifest gauge-invariance, by deferring any non-local splitting of the
perturbations until a late stage of the calculation, and by providing a
straightforward linearisation procedure around a variety of background models.

\section*{Acknowledgements}

The author would like to thank the organising committee for their invitation to
participate in the meeting and Queens' College, Cambridge, for
support in the form of a Research Fellowship.


\end{document}